\documentclass[journal,onecolumn,twoside]{IEEEtran}

\usepackage{balance}
\usepackage{algorithm}
\usepackage{algorithmic}
\usepackage{epsfig}
\usepackage{graphicx}
\usepackage{epstopdf}
\usepackage{bm,amsmath,amssymb,amsfonts,graphicx,epsfig,amsthm,color}
\usepackage{bbold,dsfont}
\usepackage{float}
\usepackage{cite}
\usepackage{amsmath,amssymb,amsfonts}
\usepackage{algorithmic}
\usepackage{graphicx}
\usepackage{enumerate}
\usepackage{textcomp}
\usepackage{threeparttable,color}
\usepackage{wrapfig}

\usepackage[hyphens]{url}
\PassOptionsToPackage{bookmarks=false}{hyperref}

\usepackage[depth=-1]{bookmark}

\newtheorem{Remark}{\textit{Remark}}
\newtheorem{Lemma}{\textit{Lemma}}
\newtheorem{Theorem}{\textit{Theorem}}

\newtheorem{Problem}{\textit{Problem}}

\newtheorem{Assumption}{\textit{Assumption}}

\newtheorem{Definition}{\textit{Definition}}
\newtheorem{Example}{\textit{Example}}

\allowdisplaybreaks

\def\BibTeX{{\rm B\kern-.05em{\sc i\kern-.025em b}\kern-.08em
    T\kern-.1667em\lower.7ex\hbox{E}\kern-.125emX}}
\allowdisplaybreaks
\begin{document}

\title{\LARGE Event-triggered Consensus Control of Heterogeneous Multi-agent Systems: Model- and Data-based Analysis}

\author{Xin Wang,~Jian Sun,~Gang Wang, and~Jie Chen,
\thanks{This work was supported in part by the National Key R$\&$D Program of China (Grant No. 2021YFB1714800), the National Natural Science Foundation of China (Grant Nos.
 62088101, 61925303, 62173034, U20B2073, and 61720106011), and in part by the Natural Science Foundation of
Chongqing (Grant No. 2021ZX4100027).

X. Wang, J. Sun, and G. Wang are with the School of Automation and the Key Laboratory of Intelligent Control and Decision of Complex System, Beijing Institute of Technology, Beijing 10081, China.~J. Sun is also with the Beijing Institute of Technology Chongqing Innovation Center, Chongqing 401120, China (e-mail: xinwang@bit.edu.cn; sunjian@bit.edu.cn; gangwang@bit.edu.cn).

J. Chen is with the Department of Control Science and Engineering, Tongji University, Shanghai 201804, China, and also with the State Key Lab of Intelligent Control and Decision of Complex Systems, School of Automation, Beijing Institute of Technology, Beijing 100081, China 	
(e-mail: chenjie@bit.edu.cn).}}
\maketitle

\begin{abstract}This article deals with model- and data-based consensus control of heterogenous leader-following multi-agent systems (MASs) under an event-triggering transmission scheme. A dynamic periodic transmission protocol is developed to significantly alleviate the transmission frequency and computational burden, where the followers can interact locally with each other approaching the dynamics of the leader. Capitalizing on a discrete-time looped-functional, a model-based consensus condition for the closed-loop MASs
is derived in form of linear matrix inequalities (LMIs), as well as a design method for obtaining the distributed controllers and event-triggering parameters. Upon
collecting noise-corrupted state-input measurements during open-loop operation, a data-driven leader-following MAS representation is presented, and employed to solve the data-driven consensus control problem without requiring any knowledge of the agents' models.
 This result is then extended to the case of  guaranteeing an $\mathcal{H}_{\infty}$ performance. A simulation example is finally given to corroborate the efficacy of the proposed distributed
event-triggering scheme in cutting off data transmissions and the
data-driven design method.
\end{abstract}


\section{Introduction}

Consensus of multi-agent systems (MASs) has gained enormous attention over the last two decades, thanks to their widespread applications in, e.g., mobile robots, sensor networks, and unmanned air vehicles \cite{Dimarogonas2012,Ma2017multisurvey,Liu2018Multi,Yang2020Eventmulti,Xiao2019,Chen2022unmanned}. Consensus control problem can be classified to leader-following and leaderless ones, depending on whether there is a leader system. So far, both cases have been widely studied, see, e.g., \cite{Liu2020multi,Qian2019multi}.
This paper focuses on the  leader-following control of  heterogeneous MASs, where agents have different dynamics.

To achieve this goal, the information interaction is required between agents via a shared network.
Considering the limited network resources (e.g., bandwidth and energy of wireless transmission nodes), intermittent transmission strategy is applicable in a digital network.
One effective approach is the event-triggering scheme (ETS), 
whose remarkable feature is that the times of transmission actions and control updates are determined by predesigned triggering conditions \cite{Chen2020}. Fruitful theoretical achievements on even-triggered consensus control of MASs are referred to a survey \cite{nowzari2019event}.
Recently, a class of ETSs known as dynamic ETS was proposed by \cite{Girard2015}. Compared to static ETS \cite{Tabuada2007} involving constant thresholds, the dynamic ETS is effective to reducing
communications by introducing a positive state-dependent dynamic threshold in the static ETS's triggering condition.
Due to this superiority,
the dynamic ETS has been gradually incorporated in MASs, such as in the cases of continuous-time \cite{Hu2020distribute} and discrete-time \cite{MISHRA2021event}, for the purpose of reducing transmission frequencies between agents. Avoiding continuous detection in the sensors, dynamic periodic distributed ETSs proposed by \cite{Borgers2018periodic,Deng2021} execute the trigger generators after the elapse of a constant time period, where
the distributed dynamic variables also need not to continuously evolve. However, these outstanding contributions are restricted to the continuous-time situations. Our work extends the dynamic periodic distributed ETS for MASs in discrete time.

On the other hand, all the above-mentioned event-triggered consensus control designs are model-based, in the sense that they require complete knowledge of all agents for the controller design and implementation.
	Nevertheless, obtaining an accurate model of a real-world system can be computationally expensive, and
the obtained models may be too complex for classic control methods to be employed \cite{Astrom1973article}.
Removing the dependency on explicit models of MASs in consensus control,
\emph{data-driven} control performs controls directly from measured data without
requiring any steps of identifying real systems, see \cite{WILLEMS2005,persis2020,Coulson2019,allibhoy2020data,berberich2019a,Liu2021data}. For example, data-driven distributed protocols achieving synchronization of MASs have been derived based on reinforcement learning techniques in \cite{Abouheaf2014reinforce,Kiumarsi2017Qlearn,Xie2021reinforcement}. However, these methods require a large number of measurements and incur high computational overhead. In \cite{jiao2021data},
an alternative approach based on Fundamental Lemma \cite{WILLEMS2005} provided an off-line data-based solution of
output regulation for the leader-following consensus problem, without iteration calculation.
But, there is a limitation in using \cite{jiao2021data}, where the disturbances are assumed known in design of the distributed control protocols.
It is more practical to consider the situation with unknown and bounded disturbances. 

These recent advances have motivated this paper to focus on data-driven leader-following consensus control of event-triggered discrete-time MASs with unknown heterogeneous dynamics. Firstly, we develop a discrete-time dynamic periodic distributed ETS, which is generalized from the aforementioned ETSs \cite{Hu2020distribute,MISHRA2021event,Borgers2018periodic,Deng2021}, to save transmission resources. By virtue of a discrete-time looped-functional (DLF) in \cite{wang2022disdata}, a model-based consensus condition for
leader-following MASs is deduced, as well as a model-based co-design method of distributed controllers and ETS parameters. Different from \cite{wang2022disdata,wang2022condata}, a simpler DLF without referring to the integral
terms of system states is employed in the system analysis to reduce the computation burden.
Then, inspired by \cite{berberich2020combining} for single agent, we formulate a data-based system parametrization formed by quadratic matrix inequality (QMI) for the MASs using noise-corrupted state-input measurements in the open-loop operation, where the disturbances are supposed to satisfy a QMI-formed bound.
By joining  the data-based representation and the model-based criterion, a data-based solution for obtaining the consensus controller and the triggering parameters without any prior knowledge of the MASs is established.  This result is finally extended to obtaining an  $\mathcal{H}_{\infty}$ performance guarantee on the closed-loop $\mathcal{L}_2$-gain, while considering the disturbances in the closed-loop process.

	In succinct form, the main contributions of this paper are listed as follows.
	\begin{enumerate}
		\item [\textbf{c1)}] We develop a novel discrete-time distributed ETS on the basis of periodic sampling for  leader-following MASs, where event-generators and dynamic variables are only executed after a predetermined time interval thus to moderate the computation frequency;
		\item [\textbf{c2)}] We establish a model-based consensus criterion for the MASs under the proposed ETS using a tailored DLF, along with a model-based design method for obtaining the distributed controllers and the ETS matrices;
        \item [\textbf{c3)}] We provide
a data-driven control method by wedding the model-based method and the data-based representation of the MASs, which is extended to the case of achieving $\mathcal{L}_2$-gain performance.
	\end{enumerate}
	
	
	The remaining structure of the article is summarized as
follows.
	In Section \ref{Sec:preliminaries}, we formulate the data-driven consensus problem for leader-following MASs, along with a data-driven MAS representation and a novel dynamic distributed ETS.
	In Section \ref{sec:Results}, model-based and data-driven methods for obtaining the distributed controller gains and the triggering matrices are presented in Section \ref{sec:Results}, as well as an extension of achieving $\mathcal{L}_2$-gain performance.
Section \ref{sec:example} certificates the practicality of our methods employing one practical example. In the final, Section \ref{sec:conclusion} shows a conclusion.
	
	{\it Notation.}
	Throughout the full paper, we denote by $\mathbb{N}$, $\mathbb{R}^n$, and $\mathbb{R}^{n\times m}$ the set of all
non-negative
	integers, $n$-dimensional real vectors, and ${n\times m}$ real matrices, respectively. For any integers $a,b\in \mathbb{N}$,  define that $\mathbb{N}_0:=\mathbb{N}\cup \{0\}$ and
	$\mathbb{N}_{[a,b]}:=\mathbb{N}\cap [a,b]$.
	The superscripts `$-1$' and `$\top$' stand for the inverse and transpose of a matrix;
	Further, we write $P\succ 0$ ($P\succeq 0$) if $P$ is a symmetric positive (semi)definite matrix. 
We write $0$ ($I$)  for a zero (identity) matrices
	of appropriate
	dimensions. 
Symbol ${\rm diag}^N_{i=1}\{q_i\}$ represents (block)diagonal matrix with $q_1,\ldots,q_N$ on its main diagonal. $\mathbf{1}_N$ $(\mathbf{I}_N)$ denotes a column vector whose elements are $1$ ($I$), and	 `$\ast$' represents the symmetric term in (block)symmetric matrices.
	${\rm Sym}\{P\}$ is the sum of $P^{T}$ and $P$. The space of square-integrable vector functions over $[0,~\infty]$ is given by $\mathcal{L}_2[0,~\infty]$ , and for $\varpi(t)\in \mathcal{L}_2[0,~\infty]$ its norm is given by $\|\varpi(t)\|_{\mathcal{L}_2}=[\int_{0}^\infty \varpi^T(t)\varpi(t)dt ]^{1/2}$. Finally,
	$\|\cdot\|$ denotes the Euclidean norm of a vector.

\section{Problem Formulation}\label{Sec:preliminaries}
	\subsection{Description of MASs}

This paper considers MASs consisting of one leader and $N$ followers.
A directed graph $\mathcal{G}:=\{\mathcal{V},\mathcal{E},\mathcal{C}\}$ is used to represent the communication topology among the agents, where $\mathcal{V}:=\{v_0, v_1, v_2, \ldots,v_N\}$ is the set of nodes, and $\mathcal{E} \subseteq \mathcal{V}\times \mathcal{V}$ represents the set of edges. The matrix $\mathcal{C}:=[c_{ij}]\in \mathbb{R}^{(N+1)\times (N+1)}$ is the adjacency of $\mathcal{G}$, constructed by setting $c_{ij}=1$ if node $v_i$ can receive information from node $v_j$ via communication channels and $c_{ij}=0$, otherwise.  Self-loops
are not taken into consideration, i.e., $c_{ii}=0$ for all $i\in\mathbb{N}_{[0,N]}$. The graph $\mathcal{G}$ is said to have a spanning tree, if there is a root node, and there exists a directed path from the root node to each other node. The neighbor set of node $i$ is denoted by $\mathcal{N}:=\{j\in \mathbb{N}_{[0,N]}|j\neq i, c_{ij}=1\}$.

We index the leader with $0$ and the follower with $1,\cdots,N$.  Their dynamics are modeled by the following discrete-time linear time-invariant systems
\begin{equation}\label{sec1:sys}
		x_i(t+1)=A_i x_i(t)+B_i u_i(t), ~t\in \mathbb{N},~i\in\mathbb{N}_{[0,N]},
	\end{equation}
where $x_i(t)\in \mathbb{R}^{n}$ denotes the state vector of agent $i$, $u_i(t)\in \mathbb{R}^{m}$ is the control input of the agent, and $A_i\in\mathbb{R}^{n\times n}$ and $B_i\in\mathbb{R}^{n\times m}$ are constant system matrices.
The MAS in \eqref{sec1:sys} is  heterogeneous, since dynamics ($A_i, B_i$) of the $N+1$ agents are different. 

Let us define $\varepsilon_i(t):=x_i(t)-x_0(t)$ as the leader-follower errors. 
Upon collecting all the errors along with the state of the leader to form $\varepsilon(t):=[\varepsilon_1^{\top}(t)~\cdots~\varepsilon_N^{\top}(t)~x_0^{\top}(t)]^{\top}$, and similarly for the inputs  $u(t):=[u_1^{\top}(t)~\cdots~u_N^{\top}(t)~ u_0^{\top}(t)]^{\top}$, we have the following entire system expression
\begin{equation}\label{sec1:sys:entire}
	\varepsilon(t+1)=A \varepsilon(t)+Bu(t), ~t\in \mathbb{N}
	\end{equation}
where
\begin{align*}
		A&:=\left[ \begin{array}{ccccc}
			{\rm diag}^N_{i=1}\{A_i\} &\mathbf{I}_N \cdot {\rm diag}^N_{i=1}\{A_{i}-A_0\}\\
			0&A_0 \\
		\end{array} \right],~
		B:=\left[ \begin{array}{ccccc}
			{\rm diag}^N_{i=1}\{B_i\}&\mathbf{I}_N \cdot (-B_0) \\
			0&B_0 \\
		\end{array} \right].
	\end{align*}

In contrast to the existing works (e.g., \cite{{Hu2020distribute,MISHRA2021event,Borgers2018periodic,Deng2021}}), this paper focuses on a more challenging situation, where the system matrices $A_i$ and $B_i$ are all assumed \emph{unknown}.
Our objective is to design a distributed control strategy for unknown MASs \eqref{sec1:sys} with intermittent communication to ensure the
leader-following consensus asymptotically as $\lim_{t\rightarrow \infty}{ (x_i(t)-x_0(t))}=0$, $\forall
i\in\mathbb{N}_{[0,N]}$.
Note that the consensus of MASs \eqref{sec1:sys} can be converted to the stability issue of system \eqref{sec1:sys:entire}, which guarantees $\lim_{t\rightarrow \infty}\varepsilon(t)=0$.

\subsection{Data-driven representation for MASs}\label{subsection:data}
A main challenge in achieving consensus control of MASs is the inability to use the system matrices. Inspired by \cite{berberich2020combining} that deals with single systems, we provide a data-driven parametrization for linear discrete-time MASs.
Suppose that a set of data $\{\{x_i(T)\}^{\rho}_{T=0},\{u_i(T)\}^{\rho-1}_{T=0}\}$ $(T\in \mathbb{N},~\rho\in \mathbb{N}_{[1,\infty)})$ satisfy the following dynamics
\begin{equation}\label{sec1:sys:perturbed}
		x_i(T+1)=A_i x_i(T)+B_i u_i(T)+D_i w_i(T) 
	\end{equation}
are available, where matrix $D_i \in \mathbb{R}^{n\times n_{w}}$ is known and assumed to have full column rank, which can model the influence of the disturbance on the subsystem.

Imitating the expression in \eqref{sec1:sys:entire}, we rewrite system \eqref{sec1:sys:perturbed} as
\begin{equation}\label{sec1:sys:perturbed:entire}
		\varepsilon(T+1)=A \varepsilon(T)+Bu(T)+D w(T),
	\end{equation}
where $w(T):=[w_1^{\top}(T)~\cdots ~w_N^{\top}(T)~ w_0^{\top}(T)]^{\top} \in \mathbb{R}^{(N+1)n_w}$ and
\begin{equation*}
D:=\left[ \begin{array}{ccccc}
{\rm diag}^N_{i=1}\{D_{i}\} &\mathbf{I}_N \cdot (-D_{0})\\
			0&D_{0} \\
		\end{array} \right].
\end{equation*}

There is no limitation on assuming that the measurements  $\{\{x_i(T)\}^{\rho}_{T=0},\{u_i(T)\}^{\rho-1}_{T=0}\}$ are available,
because the input $\{u_i(T)\}$ and the state $\{x_i(T)\}$ of each agent are buffered at the actuator and the sensor, respectively, during the open-loop operation.
	Here, the disturbance sequence $\{w_i(T)\}^{\rho-1}_{T=0}$  is {\it unknown}, where $w_i(T)$ covers the noise that corrupts the collected data, but assumed to satisfy the following bound by defining
\begin{equation*}
W:=\left[\begin{array}{cccc}
w(0)&w(1)&\cdots &w(\rho-1)
\end{array}
\right].
\end{equation*}
   \begin{Assumption}[\emph {Noise bound}]\label{Ass:disturbance}
		{The noise sequence $\{w_i(T)\}^{\rho-1}_{T=0}$ $(i\in \mathbb{N}_{[0,N]})$ gathered in the matrix $W$ belongs to the set
		\begin{align*}
			\mathcal{W}=\bigg\{W\in\mathbb{R}^{ (N+1)n_w\times\rho} \Big |
			\left[ \begin{array}{cc}W^\top \\I \\\end{array} \right]^\top
			Q_d
			\left[ \begin{array}{cc}W^\top \\I \\\end{array}  \right]\succeq0 \bigg\},
		\end{align*}
		where $Q_d$ is a known symmetric matrix admitting an LMI representation satisfying $\left[\begin{array}{cc}I \\0 \\\end{array} \right]^\top Q_d\left[\begin{array}{cc}I \\0 \\\end{array} \right]\prec 0$.}
\end{Assumption}
\begin{Remark}\label{remark:noise}
{Assumption \ref{Ass:disturbance} is suitable for MASs, which is an extension of \cite[Assumption 2]{berberich2020combining} that only considers single systems. In this article, we suppose that the disturbance sequence $\{w(T)\}^{\rho-1}_{T=0}$ satisfies a pointwise-in-time norm bound $\|w(T)\|\leq \bar{w}$ with $\bar{w}>0$ for all $T \in \mathbb{N}$. Then, a special case of $Q_d$ is described as
\begin{align}\label{multiplier:diagonal}
			Q_d=
			\left[ \begin{array}{cc}-{\rm diag}^\rho_{i=1}\{q_i\} & 0\\0 & \sum_{i=1}^\rho q_i \bar{w}^2 I\\\end{array} \right]
			, ~q_i\geq0.
		\end{align}
The multiplier \eqref{multiplier:diagonal} is less conservative compared to the ones used by \cite{wildhagen2021datadriven,wang2022disdata}, since it contains free scalars $q_i$ for each data point, whereas only one scalar is included in \cite{wildhagen2021datadriven,wang2022disdata}.
}\end{Remark}

	Employing the available state-input measurements $\{x_i(T)\}^{\rho}_{T=0}$, $\{u_i(T)\}^{\rho-1}_{T=0}\}$, we  compute and collect
the leader-follower errors $\varepsilon_i(T)=x_i(T)-x_0(T)$ that are consistent with the system expression in
\eqref{sec1:sys:entire} for all discrete times $T\in \mathbb{N}$. Let us arrange these data to construct the matrices as follows 
	\begin{align*}
		E_+ &:=\left[\begin{array}{cccc}
\varepsilon(1)&\varepsilon(2)&\cdots &\varepsilon(\rho) \\
\end{array}\right],\\
		 E&:=\left[\begin{array}{cccc}
\varepsilon(0)&\varepsilon(1)&\cdots &\varepsilon(\rho-1) \\
\end{array}\right],\\
		 U&:=\left[\begin{array}{cccc}
u(0)&u(1)&\cdots &u(\rho-1) \\
\end{array}\right].
	\end{align*}
Note that the noise-corrupted measurements have covered a set of pairs $[A ~B]$, since the unknown noise is bounded by Assumption \ref{Ass:disturbance}. With the above definitions, we define the set of all pairs $[A~B]$ that are consistent with the measurements and the bounded noise as the following form, based on
equation \eqref{sec1:sys:perturbed:entire} and Assumption \ref{Ass:disturbance},
	\begin{align*}
		\Sigma_{AB}:=\Big\{[A~ B] \left | E_+=AE+BU+B_wW,~ W\in \mathcal{W}\right.\Big\},
	\end{align*}

In order to guarantee stability of \eqref{sec1:sys:entire} without knowledge of system matrices (which implies leader-follower consensus of \eqref{sec1:sys}), we need to achieve a
stability criterion for all $[A~ B] \in \Sigma_{AB}$. For this purpose, a data-based representation of $[A~ B]$ expressed as a QMI is detailed as follows.

\begin{Lemma}[{\emph {Data-driven system representation}}] \label{Lemma:system:data}
		{The set $\Sigma_{AB}$ is equivalent to
		\begin{align*}
			 \Sigma_{AB}=\bigg\{[A~B] \Big |
			\left[\begin{array}{cc}[A~B]^\top \\I \\\end{array}\right]^\top
		\Theta_{AB}
			\left[\begin{array}{cc}[A~B]^\top \\I \\\end{array}\right]\succeq0
			\bigg\},
		\end{align*}
		where  $\Theta_{AB}:=
		\left[\begin{array}{cc}-E & 0 \\ -U & 0 \\ \hline E_+ & B_w \\\end{array}\right]
		Q_d
		\left[\begin{array}{cc}-E & 0 \\ -U & 0 \\ \hline E_+ & B_w \\\end{array}\right]^\top $.
	}\end{Lemma}
	
\begin{figure}
		\centering
		 \includegraphics[scale=0.6]{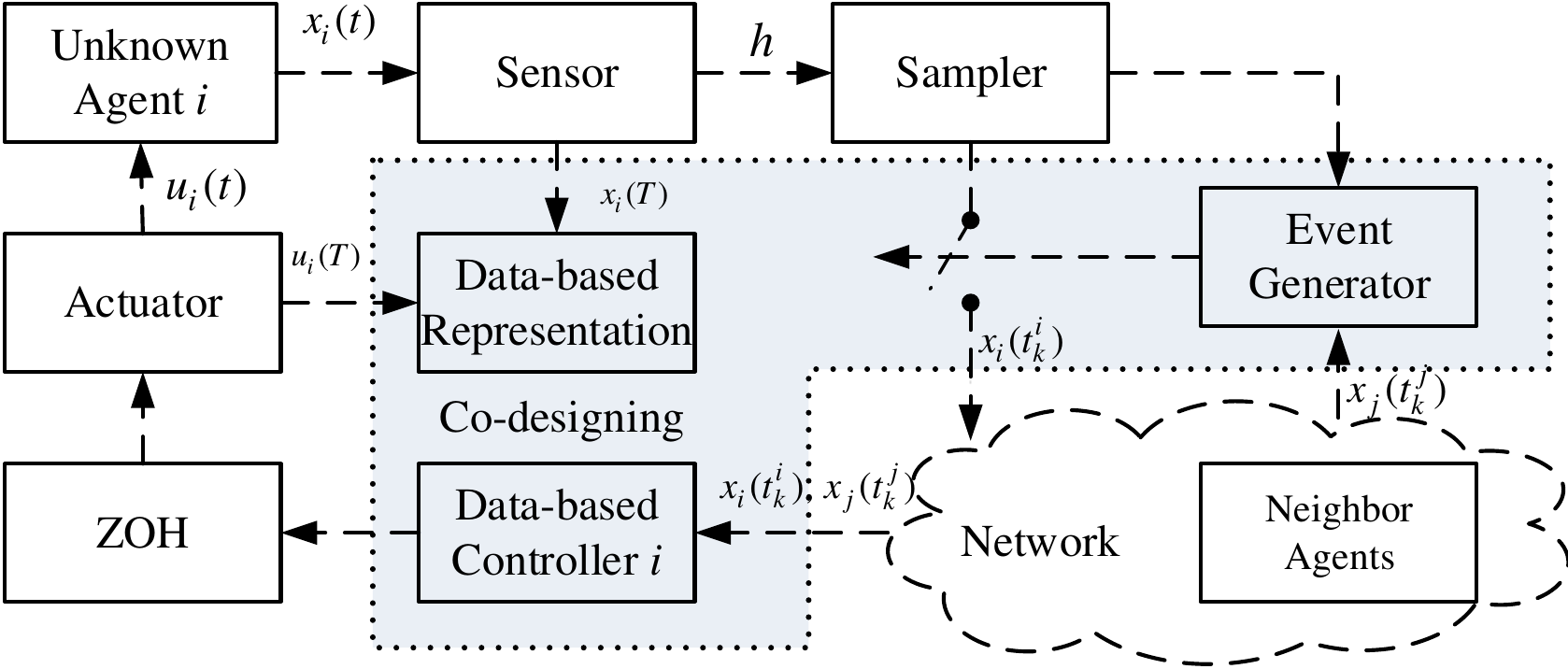}
		\caption{Data-driven networked control under event-triggering scheme.}
		\label{FIG:structure:event}
	\end{figure}

	Lemma \ref{Lemma:system:data} provides a purely data-driven parametrization of the \emph{unknown} system \eqref{sec1:sys:entire} using only data $E$, $E_+$, and $U$. Note that such data are collected from the perturbed system \eqref{sec1:sys:perturbed:entire}, while we analyze the stability of the unperturbed system \eqref{sec1:sys:entire}.
Introducing the disturbance here is to interpret the possible noise in the pre-collected state-input data, rather than to explain the disturbance affecting the system dynamics. We present the results on obtaining a performance guarantee of the closed-loop $\mathcal{L}_2$-gain in Section \ref{sec:distrubance}.
Also note that, in Fig. \ref{FIG:structure:event}, such data are collected off-line in an open-loop experiment.
By contrast, the leader-follower consensus is realized in a closed-loop experiment, which is independent of the open-loop sampling.
In the next subsection, a  distributed control strategy with an event-triggering transmission scheme is proposed to ensure an agreement of the MASs.

\subsection{Distributed event-triggered control }\label{subsec:event}

It is assumed that during the closed-loop operation, the state of each agent is periodically sampled with a common period $h$ in a synchronous manner, where $1\leq \underline{h} \leq h \leq \bar{h}$ with given bounds $\underline{h}$, $\bar{h} \in \mathbb{N}$. But, the sampled state is transmitted to local or neighbors' controllers in an asynchronous way, which is named event-triggering scheme, and it will be designed later. Denoting the $k$th transmitted instant of agent $i$ by $t_k^i$,
we consider the following linear feedback law for system \eqref{sec1:sys} during $t\in \mathbb{N}_{[t_k^i, t_{k+1}^i-1]}$
\begin{equation}\label{sec1:feedback}
u_i(t)=\left\{\begin{array}{lll}
K_{i}x_i(t_k^i), &&i=0\\
\sum_{j \in \mathcal{N}} K_{ij}e_{ij}(t_k^i),&&i>0,
\end{array}
\right.
	\end{equation}
	where $k'(t):={\rm arg} \min_{l\in \mathbb{N}:t\geq t_l^j} \{t-t_l^j\}$,
	 and therefore, for each $t\in \mathbb{N}_{[t_k^i, t_{k+1}^i-1]}$, $t_{k'(t)}^i$ is the last transmitted time of agent $i$; $e_{ij}(t_k^i):=x_i(t_k^i)- x_j(t_{k'(t)}^j)$ is the state measurement error of agents $i$ and $j$; and,
	  the consensus controller gain matrices $K_{0}$, $K_{ij}\in\mathbb{R}^{m\times n}$ are designed in Section \ref{sec:Results}.
Note that in \eqref{sec1:feedback} the leader has only access to its own sampled states,  
and the control inputs of the followers only contain errors with respect to their neighbors.
Especially,
a nonzero coupling matrix $K_{ij}$ implies that there exists a communication channel through which controller $i$ can utilize
 $e_{ij}(t_k^i)$, otherwise $K_{ij}=0$.

In our distributed control, each follower agent updates its own control input at transmitted times by capitalizing on all information locally available as well as received from its neighboring agents.
	Specifically, for all followers $i>0$, the error of the local sampled state $x_i(t_k^i)$ and the state $x_j(t_{k'(t)}^j)$ of the neighboring agent are employed, and its control input is obtained using the law in \eqref{sec1:feedback}. 
	
Next, motivated by \cite{Hu2020distribute,MISHRA2021event,Borgers2018periodic,Deng2021}, a distributed periodic event-triggering scheme is introduced to determine the transmitted instants $\{t_k^i\}$, capitalizing on the following criterion
	\begin{equation}\label{sys:judgement}
		\eta_i(\tau_v^i)+\theta_i \rho_i(\tau_v^i)<0
	\end{equation}
	with $\tau_v^i:=t_k^i+vh$ for all $v\in \mathbb{N}_{[0,m^i_k]}$, where $\theta_i>0$ is to be designed, $m^i_k=\frac{t_{k+1}^i-t_k^i}{h}-1$, and
	\begin{align*}
\rho_0(\tau_v^0)&:=\sigma_0 x_0^\top (t_k^0)\Omega_0 x_0(t_k^0)-e_0^\top (\tau_v^0)\Omega_0 e_0(\tau_v^0),\\
\rho_i(\tau_v^i)&:=\sum_{j\neq i}^N\sigma_{ij} e_{ij}^\top (t_k^i)\Omega_i e_{ij}(t_k^i)-e_i^\top (\tau_v^i)\Omega_i e_i(\tau_v^i),i>0,
	\end{align*}
	where $\Omega_i\succ0$ is a weight matrix, and $\sigma_0$, $\{\sigma_{ij}\}_{j\in\mathcal{N}}$, $\theta_i$ are parameters, both to be designed ($\sigma_{ij}=0$ when there is no transmission path from agents $i$ to $j$); $e_i(\tau_v^i):=x_i(\tau_v^i)-x_i(t_k^i)$ denotes the error of agent $i$ between the latest transmitted signal $x_i(t_k^i)$ and the current sampled signal $x_i(\tau_v^i)$; and,
	$\eta_i(\tau_v^i)$ in condition \eqref{sys:judgement} is a discrete-time variable satisfying
	\begin{align}\label{sys:dynamic}
		 \eta_i(\tau_{v+1}^{i})-\eta_i(\tau_v^{i})=-\lambda_i\eta_i(\tau_v^{i})+\rho_i(\tau_v^{i})
	\end{align}
	where $\eta_i(0)\geq0$ and $\lambda_i>0$ are given parameters. 
To sum up, the event-triggering policy can be described as
	\begin{align}\label{sys:trigger}
		&t_{k+1}^i=t_k^i+h\min_{v\in \mathbb{N}}\Big\{v>0\Big|\eta_i(\tau_v^{i})+\theta_i\rho_i(\tau_v^{i}) < 0\Big\}.
	\end{align}

    In our distributed control strategy, the current sampled state $x_i(\tau_v^{i})$
is transmitted to the local controller, and neighbors through different paths for the following agent $i>0$, as soon as the condition \eqref{sys:judgement} is met.
According to the control law \eqref{sec1:feedback}, a new control input is computed using the state $x_i(t_{k+1}^{i})$ and received neighbor's state $x_j(t_{k'(t)+1}^j)$ (only for the followers), and kept by a zero-order holder (ZOH) during $[t_{k+1}^i, t_{k+2}^i-1]$. Meanwhile, the event-triggering element is renewed using the latest transmitted measurements and waits for checking the next received one.
The following lemma is useful for deriving our results in Section \ref{sec:Results}.
	
	\begin{Lemma}[\emph {Non-negativity}]\label{lemma:nonneg.dynam}
		{For any positive definite matrix $\Omega_i\succ 0$, non-negative scalar $\eta_i(0)\ge0$, and positive constants $\lambda_i>0$, $\theta_i> 0$ satisfying $1-\lambda_i-\frac{1}{\theta_i}\geq0$.
		Then, for all $v\in \mathbb{N}_{[0,m^i_k]}$, it holds that $\eta_i(\tau_v^{i})\geq0$ under the triggering condition \eqref{sys:trigger}.
	}\end{Lemma}
	
The proof of Lemma \ref{lemma:nonneg.dynam} is similar to \cite[Lemma 3]{wang2022disdata}, which is omitted here.

	\begin{Remark}\label{generaN} {The transmission scheme \eqref{sys:trigger} can be seen as a discrete-time counterpart of the continuous dynamic event-triggering scheme in \cite{Borgers2018periodic,Deng2021}.
			Besides, 
			our event-triggering scheme subsumes the dynamic ETS proposed in \cite{MISHRA2021event} (cf. \eqref{sys:trigger} with $h=1$) as special cases.
In \eqref{sys:trigger}, since every event-generator is only executed at sampling times $\tau_v^i$, the variable does not need to successively evolve at every discrete time in the event-generator.
			Our triggering scheme is expected to further reduce
			data transmissions and computational burden when compared to \cite{MISHRA2021event}.
Note that the dynamic thresholds $\eta_i(\tau_v^{i})$ remain positive definite according to Lemma \ref{lemma:nonneg.dynam},  thus to provide less transmissions compared to the static ones (cf. \eqref{sys:trigger} with $\theta_i\rightarrow \infty$).
	}\end{Remark}

\subsection{Problem statement}\label{Problem}
Having introduced MASs with an event-triggered strategy and the data-driven system representation, the problem 
considered is described in this section. At the beginning, we put forward the following closed-loop system expression  combining the feedback control law \eqref{sec1:feedback} and the open-loop system in  \eqref{sec1:sys:entire}
\begin{equation}\label{sec1:sys:sampling:entire}
	\varepsilon(t+1)=A \varepsilon(t)+BK\varepsilon(t_k), ~t\in \mathbb{N}_{[t_k^i, t_{k+1}^i-1]}
	\end{equation}
	where $\varepsilon(t_k):=[\varepsilon_1^{\top}(t_k^1)~\cdots~\varepsilon_N^{\top}(t_k^N)~ x_0^{\top}(t_{k'(t)}^0)]^{\top}$ and $\varepsilon_i(t_k^i):=x_i(t_k^i)-x_0(t_{k'(t)}^0)$,
	\begin{align*}
		K&:=\left[ \begin{array}{ccccc}
			\sum_{j \in \mathcal{N}} K_{1j}& -K_{12} & \cdots & -K_{1N}& 0 \\
			\vdots   & \vdots   &\ddots & \vdots &\vdots\\
			-K_{N1}  & -K_{N2} & \cdots & \sum_{j \in \mathcal{N}} K_{Nj} & 0\\
            0  & 0  &\cdots & 0 & K_0\\
		\end{array} \right],
	\end{align*}
where the transmission instant $t_k^i$ is determined by the triggering law \eqref{sys:trigger}. Based on  \eqref{sec1:sys:sampling:entire},
the problem is given as follows.
\begin{Problem}[\emph {Data-driven consensus}]\label{Pro}
{Given state-input measurements $\{x_i(T)\}^{\rho}_{T=0}$, $\{u_i(T)\}^{\rho-1}_{T=0}\}$ of the MASs \eqref{sec1:sys}, and a directed  graph $\mathcal{G}$, design a control law of the form \eqref{sec1:feedback} as well as
a triggering strategy in the form of \eqref{sys:trigger} (cf. system \eqref{sec1:sys:sampling:entire}), such that, for any initial states $x_i(0)$, $\lim_{t\rightarrow \infty}{ (x_i(t)-x_0(t))}=0$, $\forall
i\in\mathbb{N}_{[0,N]}$.
}\end{Problem}

\section{Main Results}\label{sec:Results}
	This section provides a data-driven consensus control strategy for MASs \eqref{sec1:sys}, which solves Problem \ref{Pro}. In particular, two steps are taken into consideration. A model-based analysis of the MASs under the ETS is performed in Section \ref{sec:modle:stability}, based on the discrete-time looped-functional (DLF) approach in \cite{wang2022disdata}.  Next, a data-driven design strategy for  obtaining the control gains and the ETS matrices is studied in Section \ref{design:event}.
In Section \ref{sec:distrubance}, we further extend the data-driven results in Section \ref{design:event} to the case of achieving $\mathcal{L}_2$-gain performance.
Before moving on, the following lemma is required to obtain our results. 
\begin{Lemma}\label{lemma:summation}
{For any matrices $R\in\mathbb{R}^{n\times n}\succ0$, $N\in\mathbb{R}^{m\times n}$, vectors $\vartheta\in\mathbb{R}^m$, and 
a sequence $\{x(s)\}_{s=\alpha}^{\beta-1}$,
it holds for $\alpha\leq\beta \in \mathbb{N}$ that
\begin{equation*}\label{Lemma:inequality}
\begin{aligned}
-\sum \limits_{i=\alpha}^{\beta-1} y^\top(i)R y(i)
\leq (\beta-\alpha)\vartheta^{\top} M R^{-1} M^{\top}\vartheta
+{\rm Sym}\left\{\vartheta^{\top} M [x(\beta)-x(\alpha)]\right\}
\end{aligned}
\end{equation*}
with defining $y(i):=x(i+1)-x(i)$.
}\end{Lemma}

Lemma \ref{lemma:summation} is a simpler case that without considering
summation terms in the right side of the inequality in \cite[Lemma 2]{Chen2017summation}. The proof is similar to \cite[Lemma 2]{Chen2017summation}, which is not displayed here. Lemma \ref{lemma:summation} provides the basis for the following results.
	\subsection{Model-based consensus and controller design}\label{sec:modle:stability}
	\begin{Theorem}[\emph {Model-based consensus}]\label{Th1}
		{Consider the system \eqref{sec1:sys} under the triggering condition \eqref{sys:trigger} and the control law \eqref{sec1:feedback}.  Given positive scalars $\sigma_{0}$, $\sigma_{ij}$,  $\bar{h}$, $\underline{h}$, and $\lambda_i$, $\theta_i$ satisfying $1-\lambda_i-\frac{1}{\theta_i}\geq0$ for all $i\in\mathbb{N}_{[0,N]}$ and $j\in \mathcal{N}$, asymptotic consensus of the system is achieved,
and dynamic values $\eta_i(\tau_v^{i})$ converge to the origin for any $\eta_i(0)\ge0$, if there exist matrices  $R_1\succ0$, $R_2\succ0$, $P\succ0$,
		$S=S^\top$, $M_1$, $M_2$, $F$, and $\Omega_i\succ0$ for all $i\in\mathbb{N}_{[0,N]}$, satisfying the following LMIs $\forall h\in\{\underline{h},\bar{h}\}$
		\begin{align}{\label {Th1:LMI1}}
			&\left[
			\begin{array}{ccc}
				\Xi_0+h\Xi_\varsigma+\Psi+\mathcal{Q}  & hM_\varsigma\\
				\ast & -hR_\varsigma
			\end{array}
			\right]\prec0, ~\varsigma=1,2
		\end{align}
		where
		\begin{align*}
			\Xi_0&:={\rm Sym}\big\{M_1(H_1 -H_3) + M_2 (H_4 -H_1) \big\}
+H_2^\top  P H_2 -  H_1^\top  P H_1 \\
&~~~~+ (H_2 - H_1)^\top  (R_2 - R_1) (H_2 - H_1)-\left[H_3^\top ,H_{4}^\top \right] S \left[H_3^\top ,H_{4}^\top \right]^\top \\
			\Xi_1&:=(H_2-H_1)^\top  R_2(H_2-H_1)-\left[H_3^\top ,H_{4}^\top \right]S \left[H_3^\top ,H_{4}^\top \right]^\top\\
			\Xi_2&:=(H_2-H_1)^\top  R_1(H_2-H_1)+\left[H_3^\top ,H_{4}^\top \right]S \left[H_3^\top ,H_{4}^\top \right]^\top\\
			\Psi&:={\rm Sym}\big\{F(AH_1+BKH_{5}-H_{2})\big\}\\
			\mathcal{Q}&:= H_5^\top  \mathrm{\Omega}_a H_5-
\left[\begin{array}{cc}
				H_3\\
                H_{5}
			\end{array}\right]^\top
\left[\begin{array}{cc}
				\Omega_b&-\Omega_b\\
                \ast&\Omega_b
			\end{array}\right] \left[\begin{array}{cc}
				H_3\\
                H_{5}
			\end{array}\right]\\
H_\iota&:=\left[0_{n\times (\iota-1)n},~I_n,~0_{n\times (5-\iota)n} \right], ~ \iota=1,\ldots,5,~H_0:=0_{n\times7n}\\
\Omega_{ai}&:=\sigma_{i0}\Omega_i+\sum_{j \in \mathcal{N}} \sigma_{ij}\Omega_i+\sigma_{ji}\Omega_j\\
\Omega_a&:= \left[\begin{array}{ccccc}
				\Omega_{a1}&\cdots& -\sigma_{1N}\Omega_1-\sigma_{N1}\Omega_N&0\\
				\ast&\ddots&\vdots&\vdots\\
				\ast&\ast&\Omega_{aN}&0\\
\ast&\ast&\ast&\Omega_{0}\\
			\end{array}\right],~
\Omega_b:=\left[\begin{array}{ccccc}
{\rm diag}^N_{i=1}\{\Omega_{i}\} &\mathbf{I}_N \cdot {\rm diag}^N_{i=1}\{\Omega_{i}\}\\
				\ast&\sum_{i=0}^{N} \Omega_i\\
			\end{array}\right].
		\end{align*}
	}\end{Theorem}

	{\it Proof.}
Considering the intervals $[\tau_v^i, \tau_{v+1}^i-1]$ for all $v\in\mathbb{N}_{[0,m_k^i]}$, 
	we choose a functional candidate for system \eqref{sec1:sys:sampling:entire} as follows
	\begin{align}{\label {Th1:Vt}}
		V(t)=V_a(t)+V_d(t)+ t\sum_{i=1}^N[\eta_i(\tau_{v+1}^i)-\eta_i(\tau_{v}^i)]
	\end{align}
	where Lypapunov functional $V_a(t)=\varepsilon^\top (t)P\varepsilon(t)$, $P \succ 0$; the dynamic variable $\eta_i(\tau_{v}^i)$ is provided as in \eqref{sys:dynamic}; and,
	the DLF $V_d(t)$ is designed as
\begin{equation}\label{Th1:loop}
		\begin{aligned} V_d(t)=&~(t-\tau_v^i)(\tau_{v+1}^i-t)\big[x^\top ({\tau_v^i}),~x^\top ({\tau_{v+1}^i})\big] S \big[x^\top ({\tau_v^i}),~x^\top ({\tau_{v+1}^i})\big]^\top\\
			&+(\tau_{v+1}^i-t)\Bigg[\sum \limits_{s=\tau_v^i}^{t} y^\top (s)R_1y(s) -y^\top (t)R_1y(t) \Bigg]
			+(t-\tau_v^i)\Bigg[\sum \limits_{s=t}^{\tau_{v+1}^i} y^\top (s)R_2y(s) -y^\top (t)R_2y(t) \Bigg]
		\end{aligned}
	\end{equation}
	where $y(s):=x(s+1)-x(s)$, and $S=S^\top$, $R_1\succ 0$, $R_2\succ 0$.
		
	The forward difference of the functional $V(t)$ is given as
	\begin{align}{\label {Th1:Vd}}
	\Delta V(t)= \Delta V_a(t)+ \Delta V_d(t)+
\sum_{i=1}^N[\eta_i(\tau_{v+1}^i)-\eta_i(\tau_{v}^i)]
	\end{align}
	where 
	\begin{align*}
	\Delta V_a(t)&=\xi^\top (t)\Big(H_2^\top  P H_2 - H_1^\top  P H_1 \Big)\xi(t),\\
	\Delta V_{d}(t)&=\xi^\top (t)\Big[(H_2 - H_1)^\top  (R_2 - R_1) (H_2 - H_1)-\left[H_3^\top ,H_{4}^\top \right] S \left[H_3^\top ,H_{4}^\top \right]^\top\Big]\xi(t)\\
	&~~~-\sum \limits_{s=\tau_v^i}^{t-1} y^\top (s)R_1y(s)-\sum \limits_{s=t}^{\tau_{v+1}^i-1} y^\top (s)R_2y(s),
	\end{align*}
with $\xi(t):=[\varepsilon^\top (t),~\varepsilon^\top (t+1),~ \varepsilon^\top (\tau_v^i),~\varepsilon^\top (\tau_{v+1}^i),~\varepsilon^\top (t_k)]^\top $. By Lemma \ref{lemma:summation}, we have that

\begin{align}
-\sum \limits_{s=\tau_v^i}^{t-1} y^\top(i)R y(i)
&\leq \xi^\top (t)\left[(t-\tau_v^i) M_1 R_1^{-1} M_1^{\top}
+M_1(H_1 -H_3)\right]\xi (t)\\
-\sum \limits_{s=t}^{\tau_{v+1}^i-1} y^\top (s)R_2y(s) &\leq \xi^\top (t)\left[ (\tau_{v+1}^i-t)M_2 R_2^{-1} M_2^\top+ M_2 (H_4 -H_1) \right]\xi (t).
\end{align}

Through the descriptor method \cite{Fridman2010}, we have the following equation according to the system representation \eqref{sec1:sys:sampling:entire}
\begin{equation}{\label {Th1:zero}}
	\begin{aligned}
		0&=2\xi^\top (t)F \big[A \varepsilon(t)+BK\varepsilon(t_k)- \varepsilon(t+1) \big] \\
		&=2\xi^\top (t)F \big(AH_1+BKH_{7}-H_{2}\big)\xi(t).
	\end{aligned}
\end{equation}

Summing up \eqref{Th1:Vd}-\eqref{Th1:zero} gives rise to
\begin{equation}\label{theorem1:vd}
\begin{aligned}
			\Delta V(t)\leq &~
 \xi^{\top}(t)\left[(t-\tau_v^i)(\Xi_{1} + M_1 R_1^{-1} M_1^\top) +(\tau_{v+1}^i-t)(\Xi_{2} + M_2 R_2^{-1} M_2^\top) +\Xi_0+\Psi \right]\xi(t)\\
 &+\sum_{i=1}^N[\eta_i(\tau_{v+1}^i)-\eta_i(\tau_{v}^i)].
		\end{aligned}
\end{equation}

In light of the triggering condition \eqref{sys:trigger}, Lemma \ref{lemma:nonneg.dynam} asserts that $\eta_i(\tau_{v}^i)\geq$ for $\eta_i(0)\ge0$, $\Omega_i\succ 0$, and $\lambda_i>0$, $\theta_i> 0$ satisfying $1-\lambda_i-\frac{1}{\theta_i}\geq0$. Then, according to the equation
\eqref{sys:dynamic}, it holds that
\begin{equation}\label{theorem1:trigger}
\sum_{i=1}^N[\eta_i(\tau_{v+1}^i)-\eta_i(\tau_{v}^i)]
\leq \xi^{\top}(t)\mathcal{Q}\xi(t).
\end{equation}

From \eqref{theorem1:vd} and \eqref{theorem1:trigger}, the difference $\Delta V(t)$ satisfies
\begin{equation}\label{th1:vd:final}
			\Delta V(t)\leq 
 \xi^{\top}(t)\left[\frac{t-\tau_v^i}{h}\Upsilon_1(h) +\frac{\tau_{v+1}^i-t}{h}\Upsilon_2(h) \right]\xi(t)
		\end{equation}
	where $\Upsilon_\varsigma(h)=\Xi_0+h\Xi_{\varsigma}+\Psi+\mathcal{Q} +h M_\varsigma R_\varsigma^{-1} M_\varsigma^\top $, $\varsigma=1,2$.
	
	According to the Schur Complement Lemma, inequalities $\Upsilon_1(h)\prec0$ and $\Upsilon_2(h)\prec0$ are equivalent
	to the LMIs in \eqref{Th1:LMI1}, which are convex with respect to $h$. Therefore, the LMIs in \eqref{Th1:LMI1} at the vertices of $[\underline{h},\bar{h}]$ certificate $\Delta V(t)<0$ $\forall h\in[\underline{h},\bar{h}]$.
	By the DLF approach in \cite{wang2022disdata}, it holds that $\forall \varepsilon(\tau_{v}^i)\neq 0$
	\begin{equation}{\label {Th1:vj}}
\begin{aligned}	 \sum_{s=\tau_{v}^i}^{\tau_{v+1}^i-1}\Delta V(s)=&~V_a(\tau_{v+1}^i)+(h-1)\sum_{i=1}^N \eta_i(\tau_{v+1}^i)-V_a(\tau_{v}^i)
	-(h-1)\sum_{i=1}^N  \eta_i(\tau_{v}^i)<0
\end{aligned}
	\end{equation}
which implies
\begin{equation}
\begin{aligned}	 V_a(\tau_{v+1}^i)+(h-1)\sum_{i=1}^N \eta_i(\tau_{v+1}^i)<V_a(\tau_{v}^i)
	+(h-1)\sum_{i=1}^N  \eta_i(\tau_{v}^i).
\end{aligned}
	\end{equation}
	Finally, we conclude that, on the basis of $V_a(t)>0$ and $\eta_i(\tau_{v}^i)>0$, the errors of system \eqref{sec1:sys:sampling:entire} and $\eta_i(\tau_v^{i})$ converge to the origin under the triggering condition \eqref{sys:trigger} and the feedback control law \eqref{sec1:feedback}, which also implies that MASs \eqref{sec1:sys} achieve asymptotic consensus. This completes the proof.
	
\begin{Remark}[\emph {DLF}]
{Compared to \cite{wang2022disdata}, a simper DLF that only contains sampled states of the agents is constructed in \eqref{Th1:loop}, whose aim is to reduce the matrices in the resulting consensus condition
at the expense of the conservatism.   Besides, obtaining a sampling-dependent condition (cf. Theorem \ref{Th1}) is another reason for introducing the DLF \eqref{Th1:loop}. An allowable sampling interval can be searched for using LMIs in \eqref{Th1:LMI1}, which is beneficial for designing sampling-based triggering schemes and feedback controllers.
}\end{Remark}

	Theorem \ref{Th1} provides a stability condition for a given event-triggering scheme. A design method for obtaining the distributed controllers and the event-triggering parameters, can be derived based on Theorem \ref{Th1}, while guaranteeing the consensus.
 To this end, an algebraically equivalent system to
system \eqref{sec1:sys:sampling:entire} is given as follows by defining $\varepsilon_i(t)=G_iz_i(t)$
	\begin{equation}\label{Design:NCS}
		z(t+1)=G^{-1}AG z(t)+G^{-1}B K_c z(t_k),~t\in \mathbb{N}_{[t_k^i, t_{k+1}^i-1]}
	\end{equation}
	where $G_i\in \mathbb{R}^{n \times n}$ is a nonsingular matrix, $z(t):=[z_1^\top (t)~\cdots ~z_N^\top (t)~z_0^\top (t)]^\top $, $K_c:=KG$, and $G:={\rm diag}\{G_1~G_2~\cdots~G_N\}$.
Imitating Theorem \ref{Th1}, the following theoretical result is proposed.
\begin{Theorem}
		[{\emph{Model-based design}}]\label{Th2}
		{Consider the system \eqref{sec1:sys} under the triggering condition \eqref{sys:trigger} and the control law \eqref{sec1:feedback}. Given the same scalars as in Theorem \ref{Th1}, there exists a block controller gain $K$ such that
asymptotic consensus of the system is achieved, and $\eta_i(\tau_v^{i})$ tends to zero for any $\eta_i(0)\ge0$,
if there exist matrices $R_1\succ0$, $R_2\succ0$, $P\succ0$,
		$S=S^\top$, $M_1$, $M_2$, $G$, $K_c$, and $\bar{\Omega}_i\succ0$ for all $i\in\mathbb{N}_{[0,N]}$, satisfying the following LMIs $\forall h\in\{\underline{h}, \bar{h}\}$
		\begin{align}{\label {Th2:LMI1}}
			&\left[
			\begin{array}{ccc}
				\Xi_0+h\Xi_\varsigma+\bar{\Psi}+\bar{\mathcal{Q}}  & hM_\varsigma\\
				\ast & -hR_\varsigma
			\end{array}
			\right]\prec0, ~\varsigma=1,2
\end{align}
where $\bar{\Omega}_a$ and $\bar{\Omega}_b$ are defined similar to ${\Omega}_a$ and ${\Omega}_b$ in Theorem \ref{Th1} by replacing $\Omega_i$ with $\bar{\Omega}_i$, and
\begin{align*}
\mathcal{D}&:=(H_1+2 H_2)^\top,~
\bar{\Psi}:={\rm Sym}\big\{\mathcal{D}(AGH_{1}+BK_cH_{5}-GH_{2})\big\}\\
\bar{\mathcal{Q}}&:= H_5^\top  \bar{\Omega}_a H_5-
\left[\begin{array}{cc}
				H_3\\
                H_{5}
			\end{array}\right]^\top
\left[\begin{array}{cc}
				\bar{\Omega}_b&-\bar{\Omega}_b\\
                \ast&\bar{\Omega}_b
			\end{array}\right] \left[\begin{array}{cc}
				H_3\\
                H_{5}
			\end{array}\right].
\end{align*}
		Moreover, the desired block controller and triggering matrices are co-designed as $K=K_cG^{-1}$, $\Omega_a={G^{-1}}^\top\bar{\Omega}_a{G^{-1}}$, and $\Omega_b={G^{-1}}^\top\bar{\Omega}_b{G^{-1}}$.
	}\end{Theorem}

{\it Proof.}
Choose the following functional for the system \eqref{Design:NCS} by replacing $\varepsilon$ in \eqref{Th1:Vt} with $z$
\begin{equation}\label{Th1:loop}
		\begin{aligned} V_z(t)=&~z^\top (t)Pz(t)+(t-\tau_v^i)(\tau_{v+1}^i-t)\big[z^\top ({\tau_v^i}),z^\top ({\tau_{v+1}^i})\big] S \big[z^\top ({\tau_v^i}),z^\top ({\tau_{v+1}^i})\big]^\top+t\sum_{i=1}^N[\eta_i(\tau_{v+1}^i)-\eta_i(\tau_{v}^i)]\\
			&+(\tau_{v+1}^i-t)\Bigg[\sum \limits_{s=\tau_v^i}^{t} y_z^\top (s)R_1y_z(s) -y_z^\top (t)R_1y_z(t) \Bigg]
			+(t-\tau_v^i)\Bigg[\sum \limits_{s=t}^{\tau_{v+1}^i} y_z^\top (s)R_2y_z(s) -y_z^\top (t)R_2y_z(t) \Bigg]\\
		\end{aligned}
	\end{equation}
	where $y_z(s):=z(s+1)-z(s)$.

Since \eqref{theorem1:trigger}, 
the following inequality holds with $\varepsilon_i(t)=G_iz_i(t)$ and $\xi^{\top}(t)\mathcal{Q}\xi(t)=\xi_z^{\top}(t)\bar{\mathcal{Q}}\xi_z(t)$
\begin{equation}\label{th2:trigger}
\sum_{i=1}^N[\eta_i(\tau_{v+1}^i)-\eta_i(\tau_{v}^i)]
\leq \xi_z^{\top}(t)\bar{\mathcal{Q}}\xi_z(t)
\end{equation}
where $\xi_z(t):=[z^\top (t), ~z^\top (t+1),~ z^\top (\tau_v^i),~ z^\top (\tau_{v+1}^i), ~z^\top (t_k)]^\top$.

It can be deduced by imitating \eqref{th1:vd:final} that
	\begin{equation}\label{th2:final}
		\Delta V_z(t)\leq\xi_z^{\top}(t) \left[ \frac{t-\tau_j}{h}\bar{\Upsilon}_1(h)
		+ \frac{\tau_{j+1}-t}{h}\bar{\Upsilon}_2(h)  \right] \xi_z(t)
	\end{equation}
	where  $\bar{\Upsilon}_\varsigma(h):=\Xi_0+h\Xi_{\varsigma}+\bar{\Psi}+\bar{\mathcal{Q}} +h M_\varsigma R_\varsigma^{-1} M_\varsigma^\top, ~\varsigma=1,2.$
By Schur Complement Lemma, inequalities $\bar{\Upsilon}_1(h)\prec0$ and $\bar{\Upsilon}_2(h)\prec0$ are equivalent to the LMIs in \eqref{Th2:LMI1}. Similar to Theorem \ref{Th1}, MASs \eqref{sec1:sys} achieve asymptotic consensus under the triggering condition \eqref{sys:trigger} and the feedback control law \eqref{sec1:feedback}, with the desired $K=K_cG^{-1}$, since
	system \eqref{Design:NCS} exhibits the same dynamic behavior and stability properties as \eqref{sec1:sys:sampling:entire}, which completes the proof. 	

\subsection{Data-driven consensus and controller design}\label{design:event}
	We are now ready to provide a data-driven solution for consensus and controller design of the system \eqref{sec1:sys} with \emph{unknown} matrix pair $[A~B]$ under the triggering condition \eqref{sys:trigger} and the feedback control law \eqref{sec1:feedback}.
	The core idea, inspired by \cite{Berberich2020,wildhagen2021datadriven}, is to replace the matrix pair $[A~B]$ in Theorem \ref{Th2} with a data-driven system expression using the measurements $\{x_i(T)\}^{\rho}_{T=0}$, $\{u_i(T)\}^{\rho-1}_{T=0}\}$.
	Following this line, a data-based design method guaranteeing the consensus is obtained on the basis of  Lemma \ref{Lemma:system:data} and Theorem \ref{Th2}.
\begin{Theorem}
		[\emph{Data-driven consensus and design}]\label{Th3}
		{Consider the system \eqref{sec1:sys} under the triggering condition \eqref{sys:trigger} and the control law \eqref{sec1:feedback}. Given the same scalars as in Theorem \ref{Th1},
there exists a block controller gain $K$ such that asymptotic consensus of the system is achieved for any $[A ~B]\in \Sigma_{AB}$, and $\eta_i(\tau_v^{i})$ tends to zero for any $\eta_i(0)\ge0$,
if there exist matrices $R_1\succ0$, $R_2\succ0$, $P\succ0$,
		$S=S^\top$, $M_1$, $M_2$, $G$, $K_c$, and $\bar{\Omega}_i\succ0$ for all $i\in\mathbb{N}_{[0,N]}$, satisfying LMIs $\forall h\in\{\underline{h}, \bar{h}\}$, $\varsigma=1,2$,
		\begin{align}{\label {Th3:LMI1}}
			&\left[
			\begin{array}{ccc}
				\mathcal{T}_1& \mathcal{F}+\mathcal{T}_2& 0\\
				\ast &  \Xi_0+h\Xi_\varsigma+\bar{\Psi}+\bar{\mathcal{Q}}+\mathcal{T}_3  & hM_\varsigma\\
				\ast &  \ast & -hR_\varsigma
			\end{array}
			\right]\prec0,
		\end{align}
where
		\begin{align*}
			\hat{\Psi}&:={\rm Sym}\big\{\!-\mathcal{D}GH_{2}\big\},~\mathcal{F}:=\left[H_1^\top G^\top,H_{5}^\top K_c^\top  \right]^\top \\
			\mathcal{D}&:=(H_1+\epsilon H_2)^\top ,\mathcal{V}_1:=
			\left[\begin{array}{ccc}I & 0\\\end{array}\right],~
			\mathcal{V}_2:=
			\left[\begin{array}{ccc} 0& \mathcal{D}\\\end{array}\right]\\
			 \mathcal{T}_1&:=\mathcal{V}_1\Theta_{AB}\mathcal{V}_1^\top ,~
			 \mathcal{T}_2:=\mathcal{V}_1\Theta_{AB}\mathcal{V}_2^\top ,~
			 \mathcal{T}_3:=\mathcal{V}_2\Theta_{AB}\mathcal{V}_2^\top .
		\end{align*}
		Moreover, $K=K_cG^{-1}$ is the desired block controller matrix, and the triggering matrices are co-designed as $\Omega_a={G^{-1}}^\top\bar{\Omega}_a{G^{-1}}$, and $\Omega_b={G^{-1}}^\top\bar{\Omega}_b{G^{-1}}$.
	}\end{Theorem}
	
		{\it proof}.	 Restructuring $\bar{\Upsilon}_\varsigma(h)$ in \eqref{th2:final} of Theorem \ref{Th2} as follows
	\begin{align*}
		\bar{\Upsilon}_\varsigma(h):=
		 \left[\begin{array}{cc}[\mathcal{D}A~\mathcal{D}B]^{\top}\\I \\\end{array}\right]^{\top}
		\left[\begin{array}{cc}0 & \mathcal{F}\\\ast & 
\Xi_0+h\Xi_{\varsigma}+\hat{\Psi}+\bar{\mathcal{Q}}+h M_\varsigma R_\varsigma^{-1} M_\varsigma^\top
\\\end{array}\right]
		\left[\begin{array}{cc}[\mathcal{D}A~\mathcal{D}B]^{\top}\\I \\\end{array}\right].
	\end{align*}
	
	According to Lemma \ref{Lemma:system:data}, it is met for any $[A ~B]\in \Sigma_{AB}$
\begin{align}
\left[\begin{array}{cc}[A~B]^\top \\I \\\end{array}\right]^\top
		\Theta_{AB}
			\left[\begin{array}{cc}[A~B]^\top \\I \\\end{array}\right]\succeq0.
\end{align}
Then, the full-block S-procedure \cite{Sche2001} ensures $\bar{\Upsilon}_i(h)\prec0$ for any $[A ~B]\in\Sigma_{AB}$ if the following LMIs yield
	\begin{equation}\label{Th2:fullblock1}
		\left[\begin{array}{cc}0 & \mathcal{F}\\\ast & \Xi_0+h\Xi_{\varsigma}+\hat{\Psi}+\bar{\mathcal{Q}}+h M_\varsigma R_\varsigma^{-1} M_\varsigma^\top \\\end{array}\right]+
	 \left[\begin{array}{cc}\mathcal{V}_1\Theta_{AB}\mathcal{V}_1^{\top} & \mathcal{V}_1\Theta_{AB}\mathcal{V}_2^{\top}\\\ast &  \mathcal{V}_2\Theta_{AB}\mathcal{V}_2^{\top} \\\end{array}\right]
		\prec0.
	\end{equation}

	Through Schur Complement Lemma, the inequalities in \eqref{Th2:fullblock1} are equivalent to the LMIs in \eqref{Th3:LMI1}. Subsequently, we can draw the same conclusion as Theorem \ref{Th2} that MASs \eqref{sec1:sys} achieve asymptotic consensus under the triggering condition \eqref{sys:trigger} and the feedback control law \eqref{sec1:feedback}, with the desired $K=K_cG^{-1}$, for any $[A ~B]\in \Sigma_{AB}$.	 

\begin{Remark}[\emph{Data-driven design algorithm}]\label{Remark:design}
{Note that Theorem \ref{Th3} is a sufficient condition for achieving consensus of system \eqref{sec1:sys}. Any conclusion can not be reached if the LMIs in \eqref{Th3:LMI1} are not solvable. Here, we summarize the data-driven design procedure, assuming that Theorem \ref{Th3} contains feasible solutions.
\begin{enumerate}[Step 1:]
		\item
Collect offline data $\{x_i(T)\}^{\rho}_{T=0}$, $\{u_i(T)\}^{\rho-1}_{T=0}\}$ from all agents $i$, and construct the data matrices $E^+$, $E$, and $U$;
 \item Suppose that noise $\{w(T)\}^{\rho-1}_{t=0}$ is bounded as $\|w(T)\|_2\leq \bar{w}$ ($\bar{w}>0$) satisfying Assumption \ref{Ass:disturbance};
 \item  Build matrix $Q_d$ to form $\Theta_{AB}$ of Lemma \ref{Lemma:system:data};
 \item Choose proper parameters $\sigma_{0}$, $\sigma_{ij}$,  $\lambda_i$, $h$, $\theta_i$, $\eta_i(0)$, and search for feasible matrices $K_c$, $G$, and $\bar{\Omega}$ for \eqref{Th3:LMI1};
 \item Compute $K=K_cG^{-1}$, $\Omega_a={G^{-1}}^\top\bar{\Omega}_a{G^{-1}}$, and $\Omega_b={G^{-1}}^\top\bar{\Omega}_b{G^{-1}}$,  which is the controller gain and the required triggering parameter.
 \end{enumerate}
}\end{Remark}

	\subsection{Data-driven $\mathcal{H}_{\infty}$ consensus and controller design}\label{sec:distrubance}
This subsection deals with data-driven $\mathcal{H}_{\infty}$ consensus of MASs subject to disturbances, whose dynamics are given as follows
\begin{equation}\label{sys:disturbance}
		x_i(t+1)=A_i x_i(t)+B_i u_i(t)+B_{d}^id_i(t), ~t\in \mathbb{N},
	\end{equation}
where $d_i(t)\in \mathbb{R}^{n_w}$ is the external disturbance and belongs to $\mathcal{L}_2[0,\infty]$; $B_{d}^i \in \mathbb{R}^{n\times n_{d}}$ is a known constant matrix describing the disturbance. We assume that model matrices $A_i$ and $B_i$ are \emph{unknown}, but the state-input data $\{x_i(T)\}^{\rho}_{T=0}$, $\{u_i(T)\}^{\rho-1}_{T=0}\}$ collected offline are available from \eqref{sec1:sys:perturbed}, and noise sequence $\{w(T)\}^{\rho-1}_{t=0}$ satisfies Assumption \ref{Ass:disturbance}.

Under the feedback control law \eqref{sec1:feedback}, system \eqref{sys:disturbance} can be reformed as the following error equation
\begin{equation}\label{sys:sampling:entire:noise}
	\varepsilon(t+1)=A \varepsilon(t)+BK\varepsilon(t_k)+B_{d}d(t), ~t\in \mathbb{N}_{[t_k^i, t_{k+1}^i-1]},
	\end{equation}
where $d(t):=[d_1^{\top}(t)~\cdots ~d_N^{\top}(t)~ d_0^{\top}(t)]^{\top} \in \mathbb{R}^{(N+1)n_d}$ and
\begin{equation*}
B_d:=\left[ \begin{array}{ccccc}
{\rm diag}^N_{i=1}\{B_{d}^i\}& \mathbf{I}_N \cdot (-B_{d}^0)\\
			0&B_{d}^0 \\
		\end{array} \right].
\end{equation*}

The definition of $\mathcal{H}_{\infty}$ stabilization for system \eqref{sys:sampling:entire:noise} is given as follows.
\begin{Definition}\label{def:H}
{Given a scalar $\gamma>0$, the MASs  \eqref{sys:sampling:entire:noise} achieve $\mathcal{H}_{\infty}$ consensus with the disturbance attenuation $\gamma$ if the following conditions hold.
\begin{enumerate}
		\item [1)] The error system  \eqref{sys:sampling:entire:noise} with the controller \eqref{sec1:feedback} is asymptotically stable with zero disturbance $d(t)=0$;
		\item [2)] The following bounded $\mathcal{L}_2$-gain condition is satisfied under zero initial condition for all nonzero $d_i(t)\in \mathcal{L}_2[0,\infty]$
\begin{equation}
\sum_{t=0}^{+\infty} \varepsilon^\top(t)\varepsilon(t)\leq \sum_{t=0}^{+\infty} \gamma^2 d^\top(t)d(t).
\end{equation}
	\end{enumerate}
}\end{Definition}

Then, based on Theorem \ref{Th2}, we provide a data-driven co-design method for event-triggered MASs with external disturbance, such that system \eqref{sys:disturbance} achieves consensus stability and $\mathcal{H}_{\infty}$  performance.
\begin{Theorem}
[\emph{Data-driven $\mathcal{H}_{\infty}$ consensus and design}]\label{Th4}
		{Consider the system \eqref{sys:disturbance} under the triggering condition \eqref{sys:trigger} and the control law \eqref{sec1:feedback}. Given the same scalars as in Theorem \ref{Th1},
there exists a block controller gain $K$ such that $\mathcal{H}_{\infty}$ consensus of the system is achieved with a given disturbance attenuation $\gamma>0$ for any $[A ~B]\in \Sigma_{AB}$, and $\eta_i(\tau_v^{i})$ tends to zero for any $\eta_i(0)\ge0$,
if there exist matrices $R_1\succ0$, $R_2\succ0$, $P\succ0$,
		$S=S^\top$, $M_1$, $M_2$, $G$, $K_c$, and $\bar{\Omega}_i\succ0$ for all $i\in\mathbb{N}_{[0,N]}$, satisfying LMIs $\forall h\in\{\underline{h}, \bar{h}\}$, $\varsigma=1,2$,
		\begin{align}{\label {Th4:LMI1}}
			&\left[
			\begin{array}{cccc}
				\mathcal{T}_1& \mathcal{F}+\mathcal{T}_2& 0& 0 \\
				\ast &  \Xi_0+h\Xi_\varsigma+\hat{\Psi}+\bar{\mathcal{Q}}+\mathcal{T}_3  & hM_\varsigma&\mathcal{D}B_dG\\
				\ast &  \ast & -hR_\varsigma& 0\\
\ast &  \ast &\ast &  -\gamma^2 G^\top G
			\end{array}
			\right]\prec0
		\end{align}
where $\hat{\Psi}=\bar{\Psi}+H_1^{\top}G^\top G H_1$.
Moreover, $K=K_cG^{-1}$ is the desired block controller matrix, and the triggering matrices are co-designed as $\Omega_a={G^{-1}}^\top\bar{\Omega}_a{G^{-1}}$, and $\Omega_b={G^{-1}}^\top\bar{\Omega}_b{G^{-1}}$
}\end{Theorem}

{\it Proof.}
One can observe that \eqref{Th4:LMI1} ensures \eqref{Th3:LMI1} of  Theorem \ref{Th3}, which leads to condition $1)$ of Definition \ref{def:H} with $d(t)=0$. Now, we consider the case of $d(t)\neq0$.
The disturbance system model is written as
\begin{equation*}
	z(t+1)=G^{-1}A Gz(t)+G^{-1}BK_cz(t_k)+G^{-1}B_{d}Gd_z(t)
	\end{equation*}
with defining $\varepsilon(t):=Gz(t)$ and $d(t):=Gd_z(t)$.
Then, it follows from Schur complement lemma and Theorem \ref{Th2} that LMIs in \eqref{Th4:LMI1} imply
\begin{align}\label{th4:inequality:noise}
		\Delta V_z(t)+z^{\top}(t)G^\top Gz(t)-\gamma^2 d_z^{\top}(t)G^\top Gd_z(t)<0.
	\end{align}
Summing \eqref{th4:inequality:noise} from $t=0$ to $+\infty$ yields that
\begin{align}\label{th4:inequality:noise:sum}
		\sum_{t=0}^{+\infty}z^{\top}(t)G^\top Gz(t)<\sum_{t=0}^{+\infty}\gamma^2 d_z^{\top}(t)G^\top Gd_z(t)-\sum_{t=0}^{+\infty}\Delta V_z(t).
	\end{align}
Finally, with the zero initial condition and $V_z(+\infty)>0$, it holds that
$\sum_{t=0}^{+\infty}x^{\top}(t)x(t)<\sum_{t=0}^{+\infty}\gamma^2 d^{\top}(t)d(t)$, which meets the condition $1)$ of Definition \ref{def:H} since $G$ is nonsingular matrix. This completes the proof.

\begin{Remark}
{The  distributed control strategy \eqref{sec1:feedback} and ETS \eqref{sys:trigger} only require local information of the MASs, i.e., the agent's and its neighbors' sampled states. However, our design procedures (cf. Theorems \ref{Th1}-\ref{Th4}) rely on the global information of the network graph, e.g., when constructing the data-driven MAS representation in Lemma \ref{Lemma:system:data}.
In this sense, the presented data-driven control protocols are not fully distributed, which may restricts their applications. How to avoid using global information in data-driven consensus control design motivates our future research.
}\end{Remark}
	\section{Example and Simulation}\label{sec:example}
	
A set of four mass-spring-damper systems \cite{CHUNG2020104621} is employed in this section to examine the proposed data-driven event-triggered control method. All numerical computations were performed using Matlab, together with the SeDuMi toolbox  \cite{SeDuMi}.

\begin{figure}[!t]
\centering
\begin{minipage}[c]{0.48\textwidth}
\centering
\includegraphics[scale=1.0]{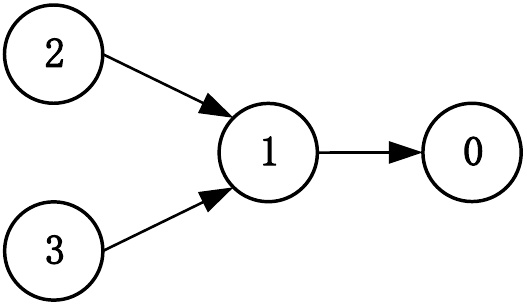}
\end{minipage}
\hspace{0.02\textwidth}
\begin{minipage}[c]{0.48\textwidth}
\centering
\includegraphics[scale=0.55]{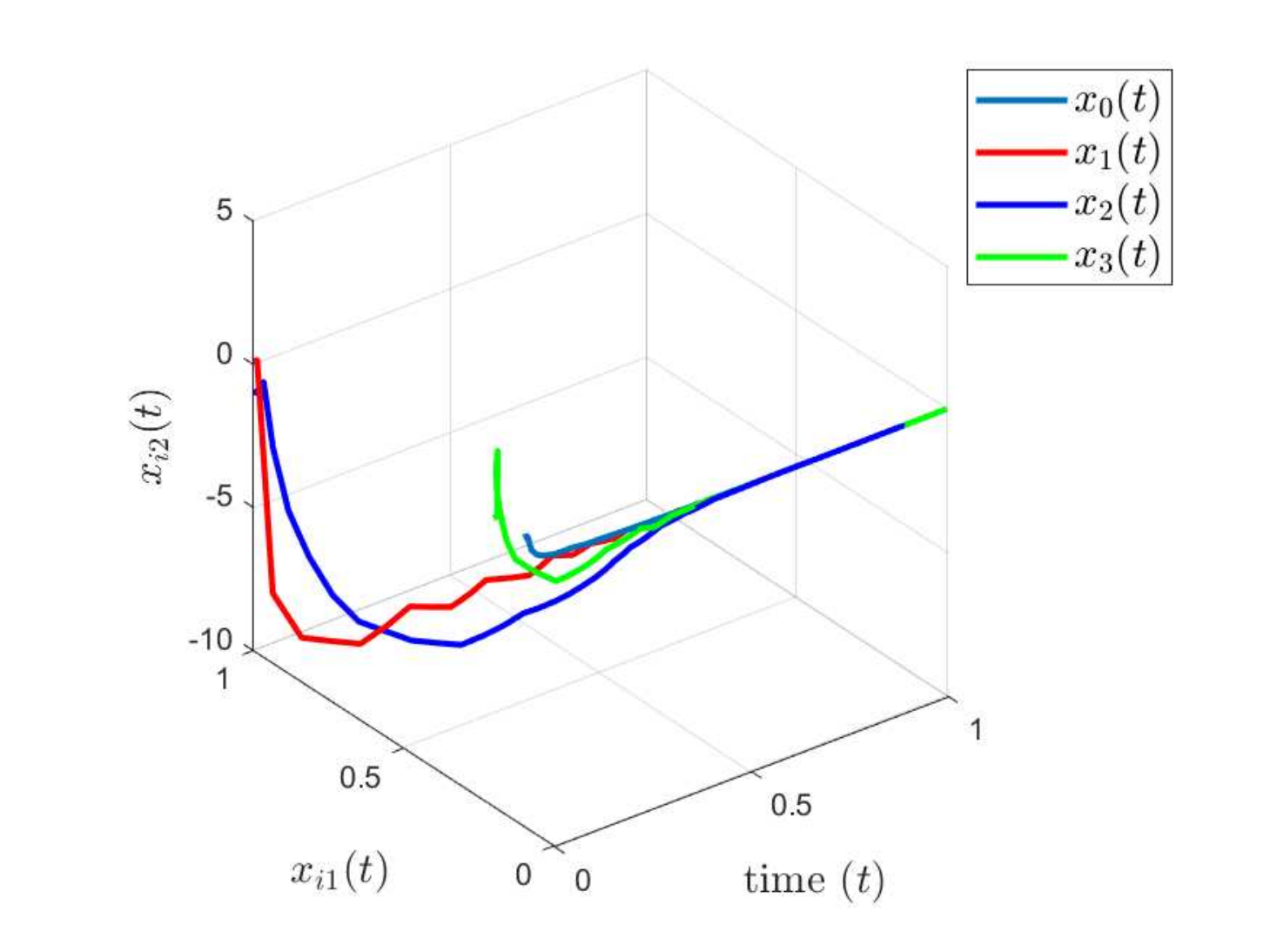}
\end{minipage}\\[3mm]
\begin{minipage}[t]{0.48\textwidth}
\centering
\caption{Communication graph.}
\label{FIG:topology}
\end{minipage}
\hspace{0.02\textwidth}
\begin{minipage}[t]{0.48\textwidth}
\centering
\caption{Trajectories of agents under the data-driven ETS.}
\label{FIG:xt}
\end{minipage}
\end{figure}

\begin{Example}{\emph {The system dynamics are given as $\dot x_i(t)=\bar{A}_ix_i(t)+\bar{B}_iu_i(t)$ with
\begin{equation*}
				\bar{A}_i=\left[
				\begin{array}{cc}
					0& 1 \\
					-\frac{f_i}{\phi_i} & -\frac{\varphi_i}{\phi_i} \\
				\end{array}
				\right],~
				\bar{B}_i=\left[
				\begin{array}{cc}
					0 \\
					\frac{1}{\phi_i}\\
				\end{array}
				\right],~ i\in\mathbb{N}_{[0,3]}
			\end{equation*}
where the state vector $x_i(t)$ consists of displacement and velocity of the mass; $u_i(t)$ is the input force; $\varphi_i$, $\phi_i$, and $f_i$ are mass, damping constant, and spring constant, respectively. Subsystems' parameters $(f_i,\phi_i,\varphi_i)^i$ are $(1,1,2)^0$, $(1,1.1,2)^1$, $(1,1.2,2)^2$, $(1,0.8,2)^3$, respectively. The interaction topology is given in Fig. \ref{FIG:topology} for a pictorial description. The adjacency matrix $\mathcal{C}$ describing the communication graph
			Upon discretization, we arrive at the discrete-time linear system as in \eqref{sec1:sys} with the matrices $A_i=e^{\bar{A_i}T_k}$ and $B_i=\int_0^{T_k}e^{\bar{A_i}s}{\bar{B_i}}ds$,
			where $T_k>0$ is the discretization interval.
			The proposed data-driven event-triggering scheme \eqref{sys:trigger} and distributed control law \eqref{sec1:feedback} are then applied to such system. The co-designed results are displayed in the following part.}}\end{Example}

	(\emph{Testing data-driven method}.)
Suppose the matrices $A_i$ and $B_i$ \emph{unknown}, but generate $\rho=40$
	measurements $\{x(T)\}^{\rho}_{T=0}$, $\{u(T)\}^{\rho-1}_{T=0}$ from the disturbed system \eqref{sec1:sys:perturbed} by setting the discretization interval as $T_k=0.01$,
where the input was generated and sampled randomly from $u(t)\in [-1,~1]$.
Besides, the collected measurements are corrupted by a disturbance satisfying $\|w(T)\|\leq 0.001$, which fulfills Assumption \ref{Ass:disturbance} as in Remark \ref{remark:noise}. 
	The matrix $B_w$ was taken as $B_w=0.01I$, which has full column rank.
	Besides, set the triggering-related parameters $\sigma_0=0.02$, $\sigma_{10}=\sigma_{21}=\sigma_{31}=0.05$, $\theta_i=5$, $\lambda_i=0.2$, and the sampling interval $h=0.01$.
	Solving the data-based LMIs of Theorem \ref{Th3} as in Remark \ref{Remark:design},
	the distributed controller gains and the event-triggering matrices were computed as
\begin{align*}
K_0&=[-683.75 ~ -71.79],
K_{10}=[-719.00 ~ -93.34],
	K_{21}=[-233.74 ~ -35.53],
	K_{31}=[-203.98 ~ -23.03]\\
	\Omega_0&=10^5\left[
	\begin{array}{cccc}
		8.6465  &  0.8485\\0.8485  &  0.0854
	\end{array}
	\right],
	\Omega_{10}=\left[
	\begin{array}{cccc}
		851.31 & 100.07\\100.07  & 13.15
	\end{array}
	\right],
	\Omega_{21}=\left[
	\begin{array}{cccc}
		485.44 &  61.70\\61.70  &  9.21
	\end{array}
	\right],
\Omega_{31}=\left[
	\begin{array}{cccc}
		716.87 &  78.29\\78.29  &  8.95
	\end{array}
	\right].
	\end{align*}


\begin{figure}[!t]
\centering
\begin{minipage}[c]{0.48\textwidth}
\centering
\includegraphics[scale=0.55]{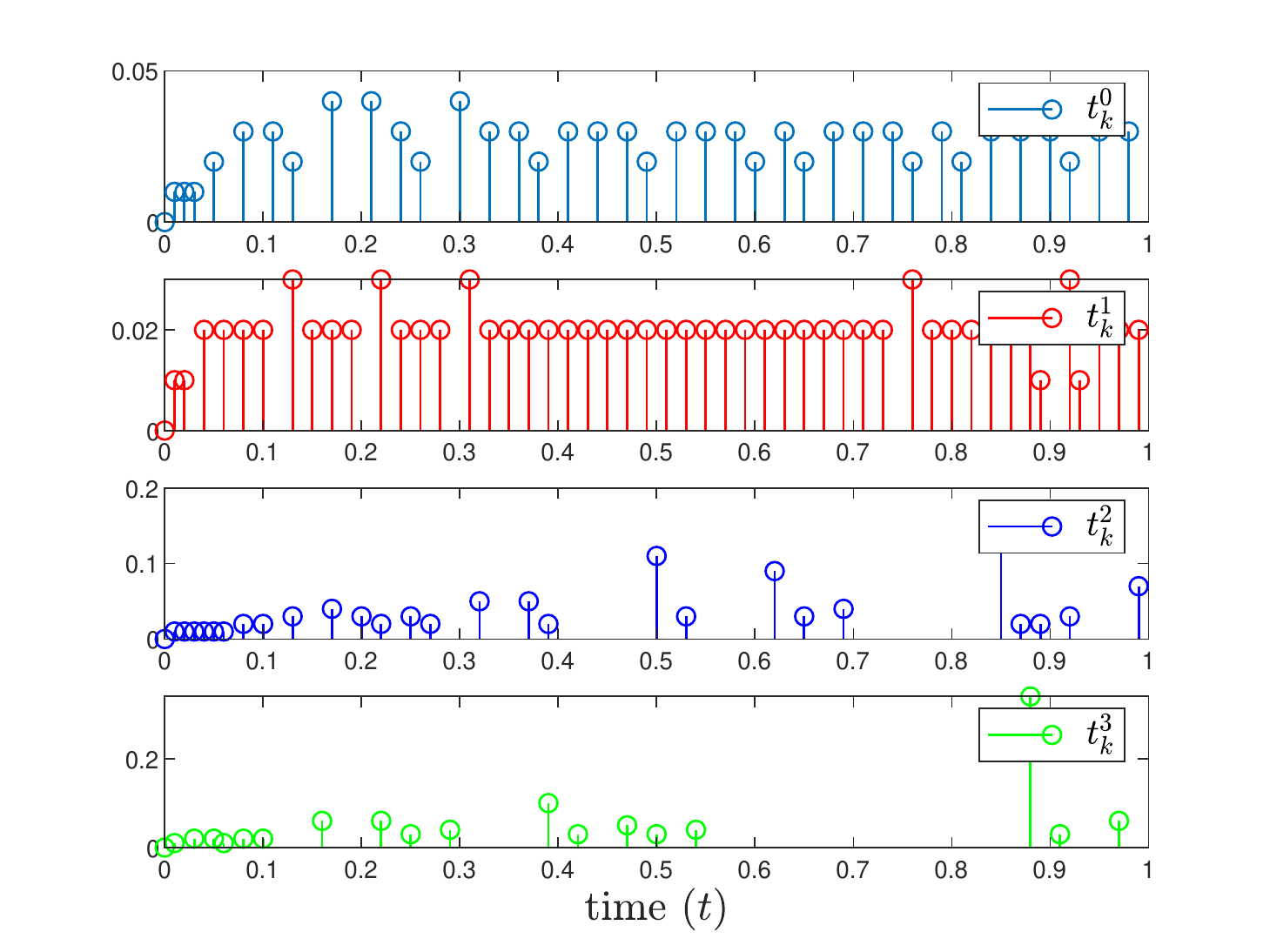}
\end{minipage}
\hspace{0.02\textwidth}
\begin{minipage}[c]{0.48\textwidth}
\centering
\includegraphics[scale=0.55]{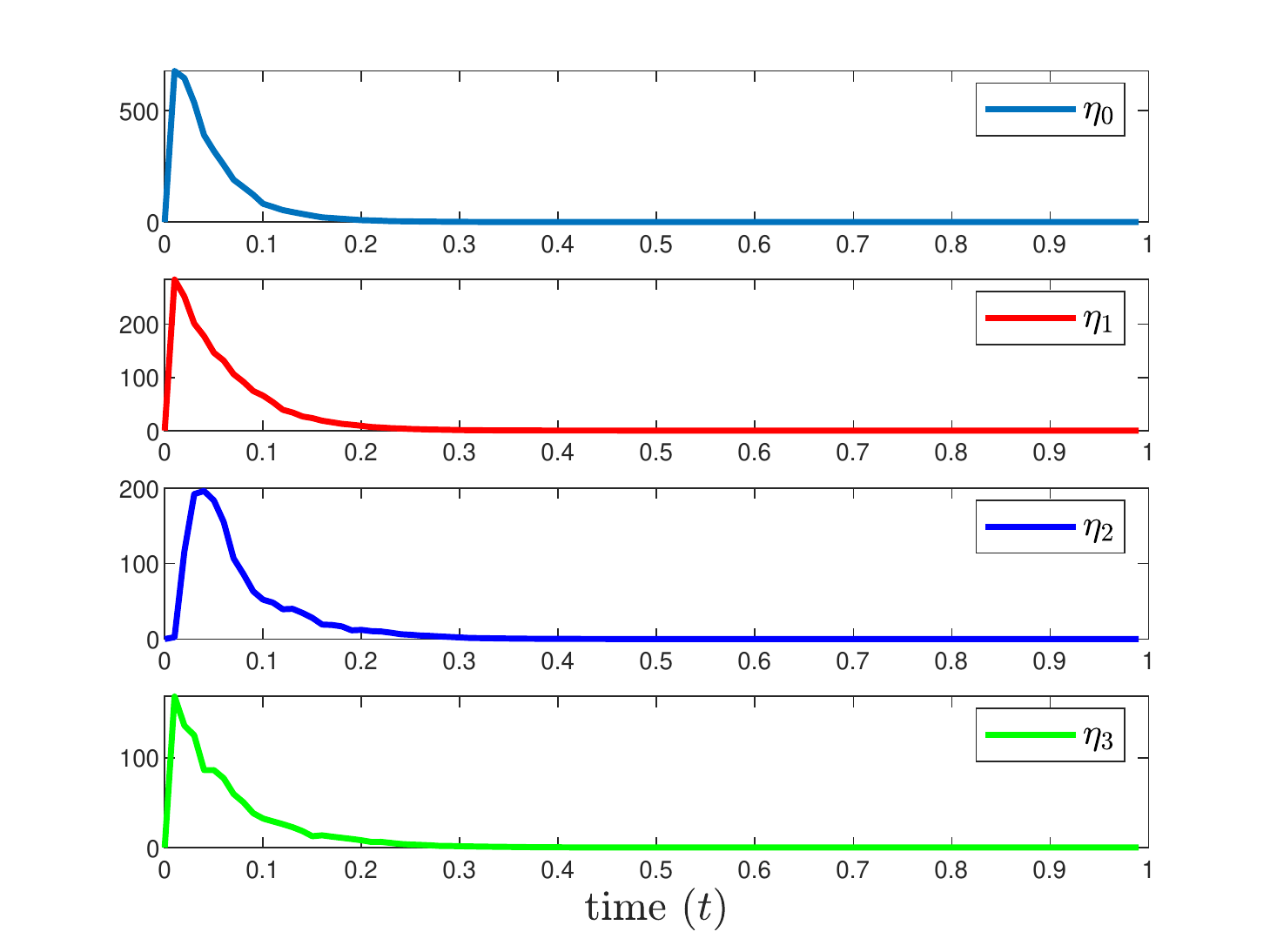}
\end{minipage}\\[3mm]
\begin{minipage}[t]{0.48\textwidth}
\centering
\caption{Broadcasting instants of agents.}
\label{FIG:tk}
\end{minipage}
\hspace{0.02\textwidth}
\begin{minipage}[t]{0.48\textwidth}
\centering
\caption{Trajectories of $\eta_i$.}
\label{FIG:dy}
\end{minipage}
\end{figure}



	
The proposed dynamic triggering scheme \eqref{sys:trigger} was numerically tested with the initial condition $x_0(0) = [0.1 ~-0.1]^{\top}$, $x_1(0) = [1~ 0.1]^{\top}$, $x_2(0) = [1~ -1]^{\top}$, $x_3(0) = [0.2~ -0.1]^{\top}$ over the time interval $[0,1]$. The trajectories of all agents
and the dynamic variables $\eta_i$ are shown in Figs. \ref{FIG:xt} and \ref{FIG:dy}, respectively.
	Obviously, all followers approach the trajectory of the leader asymptotically and the dynamic variables converge to zero,
	demonstrating the correctness of the proposed distributed data-driven triggering and control schemes.
It should also be noted that in Fig. \ref{FIG:tk} \emph{only} $37$ out of $100$ measurements for leader $0$, and $46$, $31$, $34$ for followers $1$-$3$, respectively, were broadcasted to distributed controllers and their neighbors, while a total of $100$ data were sampled for each subsystem.
	It proves that the proposed data-driven event-triggering scheme is helpful for reducing transmissions, when achieving consensus of MASs.
	
\begin{figure}[t]
		\centering
		\includegraphics[scale=0.6]{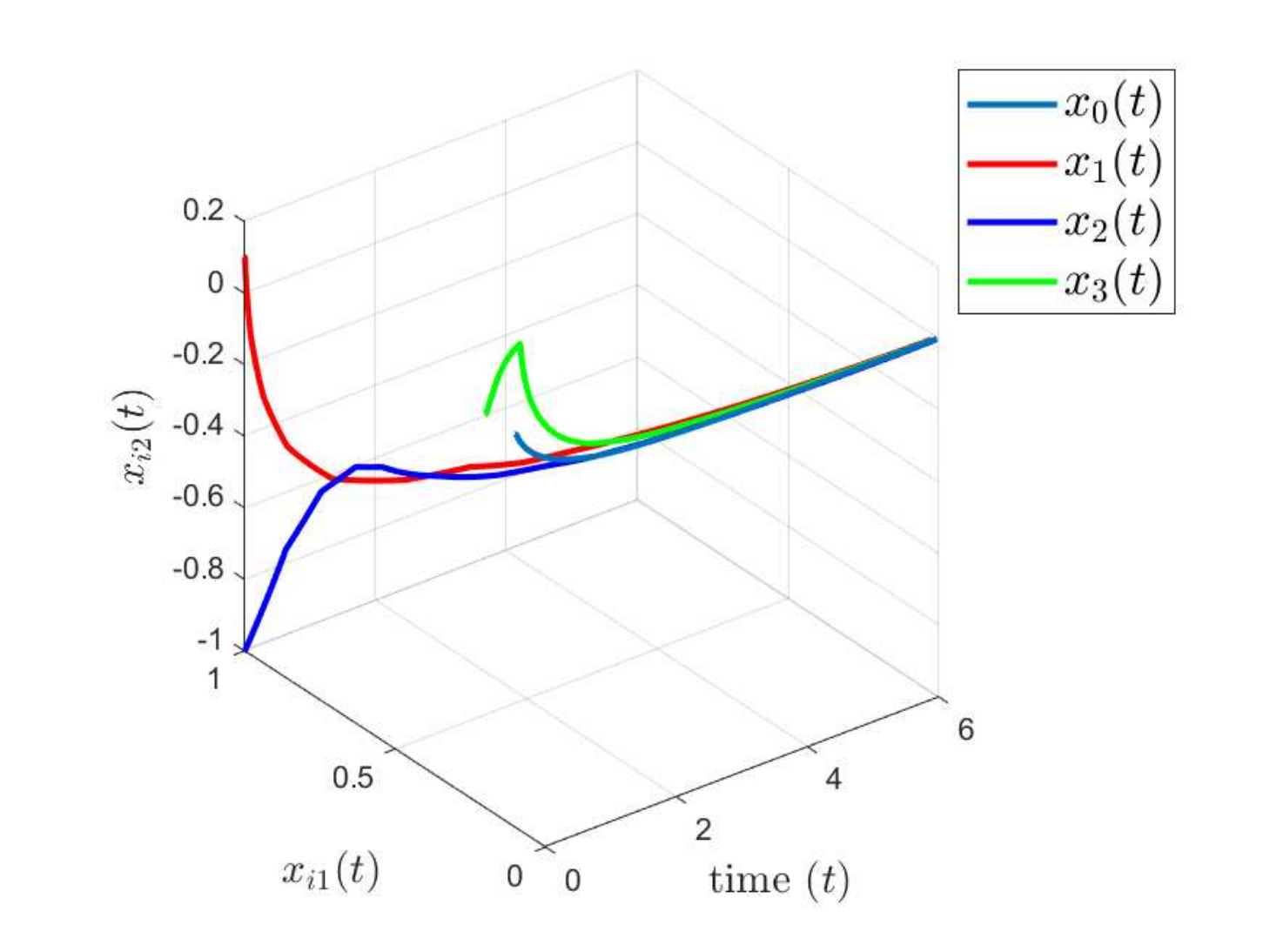}
		\caption{Trajectories of agents under the model-based ETS.}		 \label{FIG:xt:model}
	\end{figure}

(\emph{Compared to mode-based method}.) Assume the system matrices are \emph{known}. We computed the controller gains and triggering matrices, by Theorem \ref{Th2} using the same parameters as
in the data-driven case, which are
\begin{align*}
K_0&=[-0.14 ~  -0.50],
K_{10}=[-2.28 ~ -3.69],
	K_{21}=[-1.14 ~  -2.15],
	K_{31}=[-0.69 ~  -1.31]\\
	\Omega_0&=\left[
	\begin{array}{cccc}
		0.0028 &  -0.0012\\-0.0012  &  0.0036
	\end{array}
	\right],
	\Omega_{10}=\left[
	\begin{array}{cccc}
		0.0041  &  0.0031\\0.0031 &   0.0052
	\end{array}
	\right],
	\Omega_{21}=\left[
	\begin{array}{cccc}
		0.0025 &   0.0022\\0.0022  &  0.0042
	\end{array}
	\right],
\Omega_{31}=\left[
	\begin{array}{cccc}
		0.0025  &  0.0022\\0.0022  &  0.0041
	\end{array}
	\right].
\end{align*}
The state trajectories of the MASs were depicted in Fig. \ref{FIG:xt:model} under the same initial condition as in Fig. \ref{FIG:xt}. The leader-following consensus problem is also
solved by the model-based method  (cf. Theorem \ref{Th2}). Besides, compared to Fig. \ref{FIG:xt} where consensus settling time is before $t=1$, the steady-state instant of Fig. \ref{FIG:xt:model} is around  $t=4$. The main reason is that Theorem \ref{Th2} has less conservatism than Theorem \ref{Th3} (where the disturbance is introduced) at the expense of system performance. 

	\section{Concluding Remarks}\label{sec:conclusion}
	This paper considered distributed event-triggered consensus control of leader-following MASs from a data-driven vantage point.
Asymptotical consensus of the MASs under the proposed dynamic periodic distributed ETS was analyzed leveraging a novel looped-functional, as well as a model-based method for obtaining the distributed controller and ETS matrices.
Combining the data-based leader-following MAS representation and the model-based condition, a data-driven corresponding co-design approach was provided and extended to the case of achieving an $\mathcal{H}_{\infty}$  performance.
	Finally, a practical example was provided to corroborate the efficacy of the proposed event-triggering scheme in reducing transmissions, as well as the validity of our model- and data-driven co-designing methods. Our future works are centered on researching the relationship of the performance and noise-corrupted data-driven controller design.






\end{document}